\documentclass[aps]{revtex4}
\def\pochh #1#2{{(#1)\raise-4pt\hbox{$\scriptstyle#2$}}}
\def\binom#1#2{\left(\begin{array}{c}#1\\#2\end{array}\right)}
\begin {document}

\title {Nonperturbative approach to (Wiener) functional integral with $\varphi^4$
interaction}
\thanks{Presented at "PATH INTEGRALS. FROM QUANTUM INFORMATION TO COSMOLOGY",
\textit{8th International Conference,} Prague, June 6--10, 2005.}
\author{J. Boh\' a\v cik}
\email{bohacik@savba.sk} \affiliation{Institute of Physics, Slovak
Academy of Sciences, D\' ubravsk\' a cesta 9, 811 45 Bratislava,
Slovakia.}
\author{P. Pre\v snajder}
\email{presnajder@fmph.uniba.sk} \affiliation{Department of
Theoretical Physics and Physics Education, Faculty of Mathematics,
Physics and Informatics, Comenius University, Mlynsk\' a dolina
F2, 842 48 Bratislava, Slovakia.}

\begin{abstract}
We propose the another, in principe nonperturbative, method of the
evaluation of the Wiener functional integral for $\varphi^4$ term
in the action. All infinite summations in the results are proven
to be convergent. We find the "generalized" Gelfand-Yaglom
differential equation implying the functional integral in the
continuum limit.
\end{abstract}
\maketitle

We follow the definition of a functional integral as a limit of a
sequence of the finite dimensional integrals. This method has no
problems with the definition of integral measure, because for the
finite dimensional integrals the integral measure  is defined
correctly. Calculating the finite dimensional integrals we must
solve the problem of the calculation of  one dimensional integrals
\begin {equation}
I_1=\int\limits _{-\infty}^{+\infty}\;dx\;\exp\{-(a x^4+b x^2+c
x)\}\ ,
 \label {int1d}
\end {equation}
where $Re\: a>0$. The conventional perturbative approach rely on a
Taylor's decomposition of $x^4$ term with consecutive replacements
of the integration and summation order. These integrals can be
calculated, but the sum is divergent. However, $I_1=I_1(a,b,c)$ is
an entire function \textit{for any} complex values
 of $b$ and $c$, since there exist all integrals
 \[ \partial_c^n\partial_b^m I_1(a,b,c)=(-1)^{n+m}
 \int\limits _{-\infty}^{+\infty}
 \;dx\;x^{2m+n}\exp\{-(a x^4+bx^2+cx)\}\ .
 \]
Consequently, the power expansions of $I_1=I_1(a,b,c)$ in $c$ and/or
$b$ has an infinite radius of convergence (and in particular they
are uniformly convergent on any compact set of values of $c$ and/or
$b$). In what follows we shall use the power expansion in $c$:
\begin {equation}
I_1=\sum\limits _{n=0} ^{\infty} \frac{(-c)^n} {n!}\int\limits
_{-\infty}^{+\infty}\;dx\;x^n\exp\{-(a x^4+b x^2)\}\ .
 \label{sim1}
\end {equation}
The integrals appearing here can be expressed in terms of the
parabolic cylinder function $D_{\nu}(z)$, $\nu=-m-1/2$, (see, for
instance, \cite {bateman}). For $n$ odd, due to symmetry of the
integrand the integrals are zero while for  even $n=2m$ we use:
 \begin {equation}
 D_{-m-1/2}(z)\ =\
\frac{e^{-z^2/4}}{\Gamma(m+1/2)}\int_0^\infty\;dx\;
 x^{m-1/2}\exp\{-\frac{1}{2}x^2-zx\}\ .
 \label{s1}
\end {equation}

\noindent Explicitly, for the Eq.(\ref{sim1}) we have:

 \begin {equation}
 I_1=\frac{\Gamma(1/2)}{\sqrt{b}}\sum\limits _{m=0} ^{\infty}
 \frac{\xi^m}{m!}\mathcal{D}_{-m-1/2}(z)\ ,\
  \xi=\frac{c^2}{4b}\ ,\
 z=\frac{b}{\sqrt{2a}}\ ,
 \label{s2}
\end {equation}

we have used the abbreviation:
\begin {eqnarray}
 \mathcal{D}_{-m-1/2}(z)& = & z^{m+1/2}
e^{\frac{\scriptstyle z^2}{\scriptstyle 4}}\; D_{-m-1/2}(z)\
.\nonumber
\end {eqnarray}
The sum (\ref{s2}) is convergent for any values of $c$, $b$ and $a$
positive \cite{my}.

Using such expansions we shall calculate the continuum unconditional
measure Wiener functional integral:
$$\mathcal{Z} = \int [\mathcal{D}\varphi(x)]\exp (-\mathcal{S})\ ,$$ corresponding to the continuum
action with the fourth order term:
\begin {equation}
\mathcal{S} =\int \limits _0^\beta d\tau \left[c/2
\left(\frac{\partial\varphi(\tau)}{\partial\tau}\right)^2+b\varphi(\tau)^2
+a\varphi(\tau)^4\right]\ . \label{s3}
\end {equation}
Following the standard procedure, we divide the integration interval
into $N$ equal slices. We define the $N-$ dimensional integral by
the relation \cite{chai}:
\begin {equation}
\mathcal{Z}_{N}=\int\limits _{-\infty}^{+\infty} \prod \limits
_{i=1}^N\left(\frac{d\varphi_i}{\sqrt{\frac{2\pi\triangle}{c}}}\right)
\exp\left\{-\sum\limits _{i=1}^N \triangle\left[c/2
\left(\frac{\varphi_i-\varphi_{i-1}}{\triangle}\right)^2
+b\varphi_i^2+a\varphi_i^4\right]\right\} \ , \label {pcf3}
\end {equation}
where $\triangle=\beta/N$. The continuum Wiener unconditional
measure functional integral is defined by the limit:
$$\mathcal{Z} = \lim_{N\rightarrow \infty}\;\mathcal{Z}_{N}\ .$$

Applying to (\ref{pcf3}) the relation (\ref{s1}), we obtain,
\cite{my}:
\begin {equation}
\mathcal{Z}_{N} =
\left[2\pi(1+b\triangle^2/c)\right]^{-\frac{N-1}{2}}
\left[2\pi(1/2+b\triangle^2/c)\right]^{-1/2}\; \mathcal{S}_{N}
\end {equation}
with
\begin {equation}
\mathcal{S}_{N}= \sum\limits_{k_1,\cdots,k_{N-1}=0}^\infty \prod
\limits _{i=0}^N \; \left[
\frac{\left(\rho\right)^{2k_{i}}}{(2k_{i})!}
\Gamma(k_{i-1}+k_{i}+1/2)\mathcal{D}_{-k_{i-1}-k_{i}-1/2}\;(z)\right],
\label{fin1}
\end {equation}
\noindent where $k_0 = k_N = 0,$ $\rho=(1+b\triangle^2/c)^{-1}$,
$z=c(1+b\triangle^2/c)/\sqrt{2a\triangle^3}$ and the argument of the
last term $\mathcal{D}_{-k_{N-1}-1/2}(z_{last})$ is $ z_{last}
=c(1/2+b\triangle^2/c)/\sqrt{2a\triangle^3}$.
 It is worth while to stress that the
factor $\rho$ is independent of the coupling constant. The coupling
constant dependence enters into this formula only through the
argument $z$ of $\mathcal{D}$ functions.

The question of the convergence of the Eq.(\ref{fin1}) is important.
Let $a_{k_i}$ is the $k_i$ dependent part of the argument of the
product in Eq.(\ref{fin1}):
\begin {equation}
a_{k_i}= \frac{\left(\rho \right)^{2k_i}}{(2k_i)!}
\Gamma(k_{i-1}+k_i+1/2)\mathcal{D}_{-k_{i-1}-k_i-1/2}\;(z)
\Gamma(k_{i+1}+k_i+1/2)\mathcal{D}_{-k_{i+1}-k_i-1/2}\;(z)
\label{sinsum}
\end {equation}
 Using the asymptotic for parabolic cylinder functions and
Stirling formula we prove that $a_{k_i}$ approaches to zero in the
asymptotic region of $k_i$ as
$$\frac{k_i^D F^{k_i}}{k_i!\; \exp{(\sqrt{k_i})}}\; ,$$
where $D$ and $F$ are finite numbers \cite{my}. This asymptotic of
$a_{k_i}$ is sufficient for a proof of the uniform convergence of
the series not only for single $k_i$ in (\ref{fin1}), but for
arbitrary tuple  $\{k_i\}$ of indices. By the same method we can
prove the uniform convergence of the summation over two, three, ...,
$N-1$ summation indices $k_i$ in the equation (\ref{fin1}). This
important conclusion indicates, that in the formula for $Z_N$ the
$(N-1)$-tuple summation over $k_i$'s is convergent.

Following the idea of Gelfand and Yaglom the functional integral in
the continuum limit is defined by the relation
$$\lim_{N\rightarrow\infty}\, \mathcal{Z}_N = \frac{1}{\sqrt{F(\beta)}}\ ,$$
where  $F(\beta)$, for our case, is the solution of the differential
equation \cite{my}:
\begin {equation}
\frac{\partial^2}{\partial \tau^2}F(\tau)+4\frac{\partial}{\partial
\tau}F(\tau)\, \frac{\partial}{\partial \tau}\ln{S(\tau)}=
F(\tau)\left(\frac{2b}{c}-2\frac{\partial^2}{\partial
\tau^2}\ln{S(\tau)}\right)\ , \label{gyeq1}
\end {equation}
calculated at the point $\beta$ (the upper limit of the time
interval in the action (\ref{s3})). Eq.(\ref{gyeq1}) has to be
supplemented by the initial conditions: $F(0) = 1$ and $$\partial
F(\tau)/\partial \tau|_{\tau=0} = 0\ .$$ The function $S(\tau)$ is
given as the limit $N \rightarrow \infty$ of $\mathcal{S}_N$ given
by (\ref{fin1}).

The result of the summation (\ref{fin1}) (replacing the $N$
dimensional integration) is an exact relation, calculated without
any approximation. The multiple summations suppress this advantage
somewhat, therefore we shall discuss the $k_i$ summations in the
formula (\ref{fin1}). The details of this calculations are presented
in \cite{my}.

These summations can be provided by the help of the asymptotic
expansions of the parabolic cylinder functions \cite{bateman}:
\begin {equation}
e^{z^2/4}\;z^{k_i+1/2}\;D_{-k_i-1/2}(z)\;=\;\sum\limits_{j=0}^{\mathcal{J}}\;
(-1)^j \;\frac{\pochh{k_i + 1/2}{2j}}{j!\;(2z^2)^j}\ , \label{asex1}
\end {equation}
($\mathcal{J}$ is fixed finite number, $z$ is large), and the
summation relation for the generating function \cite{bateman}:
\begin {equation}
e^{x^2/4}\sum\limits_{k=0}^{\infty}\;
\frac{\pochh{\nu}{k}}{k!}\;t^k\;D_{-\nu-k}(x)\;=\;e^{(x-t)^2/4}\;D_{-\nu}\;(x-t)\
. \label{dsum1}
\end {equation}
The symbols $\pochh{k_i + 1/2}{2j}$  and $\pochh{\nu}{k}$ are the
Pochhamer's numbers defined by:
$$\pochh{\nu}{k}=\Gamma(\nu+k)/\Gamma(k)\ .
$$
In the summation over single $k_i$ in (\ref{fin1}), corresponding to
the summation over $a_{k_i}$ in (\ref{sinsum}), there appear a
product of two $D$ functions possessing as the function's subscript
the summation index. We propose to use the asymptotic expansion for
one of them. The asymptotic formula can be used in summation over
index $k_i$ provided that the summation runs over finite number of
the terms, and the index of the parabolic cylinder function, and its
argument obey the relation:
\begin {equation} k_i < N_0 < \mid z\mid\ ,
\label{cond1}
\end {equation}
$N_0$ is an upper bound of summation variables in (\ref{fin1}).

 This requirement for an application the asymptotic formula is violated
 if
the summation index approaches infinity in (\ref{fin1}). We can
overcome this problem thank to the uniform convergence of the
summations in (\ref{fin1}) proven in \cite{my}. We truncate the
summation over $k_i$ in (\ref{fin1}) but we demand that the partial
sum is close to the full infinite one. To minimize the effect of the
truncation on the final result in the continuum limit, we have to
choose the maximal value of the index $k_i$ such that the remainder
of the sum (\ref{fin1}) satisfies the inequality
\begin {equation}
\mid \sum\limits_{k_{i}=N_0+1}^\infty a_{k_i}\;\mid \; < \;
\varepsilon(N)\ , \label{rem1}
\end {equation}
with $\varepsilon(N)$ approaching zero for large $N$ so that
 $ N^3\varepsilon(N)\rightarrow 0$ in the limit $N \rightarrow \infty$.
 We have chosen the third power of $N$ in the inequality in order
 to
eliminate the influence of  remainders (\ref{rem1}) in
Gelfand-Yaglom procedure for the calculation of continuum limit. We
estimated the remainder (\ref{rem1}) and shown that both conditions
(\ref{cond1}) and (\ref{rem1}) for the application of the proposed
summation method are valid simultaneously. In this estimate we
utilized the asymptotic form of the parabolic cylinder functions
with double asymptotic properties proposed by N. Temme \cite{temme}.

The result of summations gives for (\ref{fin1}) the following
relation \cite{my}:

\begin {equation}
\mathcal{Z}_N\;=\;\left\{\prod\limits_{i=0}^{N}\;\left[2(1+b\triangle^2/c)\omega_
i\right]\right\}^{-1/2}
\;\sum\limits_{\mu=0}^{\mathcal{J}}\;\frac{(-1)^{\mu}}{\mu!\;(2z^2)^{\mu}}\;\left(N\right)_{2
\mu,\;0}\ , \label{recu1}
\end {equation}
with the symbols $\left(N\right)_{2j,\;i}$ satisfying the following
recurrence relation:
\begin {eqnarray}
\left(\alpha+1\right)_{2\mu,\;p} &=& \sum\limits_{\lambda=0}^{\mu}
\binom {\mu}{\lambda} \omega_{\alpha}^{-2(\mu-\lambda)}\;
\sum\limits^{2\lambda}_{i=max [ 0,\;p-2(\mu-\lambda)]}\;
\left(\frac{A^2}{\omega_{\alpha-1}\omega_{\alpha}}\right)^i\;
\left(\alpha\right)_{2\lambda,\;i}\;
a_p^{2(\mu-\lambda)+i}\ ,\nonumber\\
\noalign{\vskip8pt}
 \left(\alpha=1\right)_{2\mu,\;p} &=&
 \frac{a_p^{2\mu}}{\omega_0^{2\mu}},\;
 a_i^{j}\; = \;\binom{j}{i}\frac{\pochh{1/2}{j}}{\pochh{1/2}{i}}\ ,
\end {eqnarray}
where
\begin {eqnarray}
 \omega_i\; &=& \;1-\frac{A^2}{\omega_{i-1}},\nonumber\\
 A &=&\frac{\rho}{2}=\frac{1}{2(1+b\triangle^2/c)},\nonumber\\
  \omega_0 &=& 1- \frac 1
{2(1+b\triangle^2/c)}\ .
\end {eqnarray}

$\mathcal{Z}_N$ in (\ref{recu1}) is the $N-th$ approximation of the
functional integral. The continuum limit of $\mathcal{Z}_N$ we
calculate by the procedure proposed by Gelfand- Yaglom following the
analogous steps as for the calculation of functional integral for
the harmonic oscillator \cite{bkz}. This means, that for finite $N$
we construct the difference equation for (\ref{recu1}), which in the
continuum limit $N\rightarrow \infty$ is reduced to differential
equation. The method and the solutions are discussed in \cite{my},
here we present the result only.

The function $S(\tau)$ entering differential equation (\ref{gyeq1})
is calculated by:
\begin {equation}
S(\tau)=\lim_{\triangle \rightarrow 0}\,
\sum\limits_{\mu=0}^{\mathcal{J}}\;\frac{(-1)^{\mu}}{\mu!\;(2z^2\triangle^{3})^{\mu}}\;
\left(\triangle^{3\mu}(N)_{2 \mu,\;0} \right)\ ,
\end {equation}
with the expansion factor
$$\frac{1}{(2z^2\triangle^{3})^{\mu}}=\left(\frac{a \rho}{c^2}\right)^{\mu}$$
finite in such limit. The additional power $\triangle^{3\mu}$
certifies that in the relation for $(N)_{2 \mu,\;0}$ only the
leading terms in powers of $1/\triangle$ survive the continuum
limit.

Equation (\ref{gyeq1}) can be simplified by the substitution
$$F(\tau) = \frac{y(\tau)}{S^2(\tau)}$$
For $y(\tau)$ we find the equation
\begin {equation}
\frac{\partial^2}{\partial \tau^2}y(\tau) =
y(\tau)\left[\frac{2b}{c} + \left(\frac{\partial}{\partial
\tau}\ln{S^2(\tau)}\right)^2 \right]\ , \label{gyeq3}
\end {equation}
accompanied  by the  initial conditions
\begin {eqnarray}
y(0) &=& S^2(0), \label{gyeq4} \\ \frac{\partial}{\partial \tau}y(0)
&=& \frac{\partial}{\partial
\tau}S^2(\tau)|\raise-4pt\hbox{$\scriptstyle{\tau=0}$}.\nonumber
\end{eqnarray}

The evaluation of function $S(\tau)$ is not yet finished. We present
the result, related to the usual perturbative expansion, to the
lowest order terms in power expansion in the variable $a$. Up to the
term linear in coupling constant $a$ we have found \cite{my}:
$$S^{(1)}(\tau)=\lim_{\triangle \rightarrow 0}\sum\limits_{\mu=0}^{1}\;
\frac{(-1)^{\mu}}{\mu!\;(2z^2)^{\mu}}\;\left(N\right)_{2 \mu,\;0} =
1 -\frac{\pochh{1/2}{2}}{2}\; \frac{a}{4c^2\gamma^3}\;
\left\{\tanh(\tau\gamma) + \tau\gamma\left[
3\tanh^2(\tau\gamma)-1\right]\right\}\ ,$$ where $ \gamma
=\sqrt{2b/c}$.

We see, that for $b>0$ in the harmonic oscillator limit, (i.e. $a=0$
implying $S^{(1)}(\tau)\equiv 1$), Eqs. (\ref{gyeq1}, \ref{gyeq3})
reduce to Gelfand- Yaglom equation for the harmonic oscillator.  We
find that in lowest nontrivial order in $a$ the Eq. (\ref{gyeq3}) is
of the modified Bessel type one. Then the functional integral can be
expressed in term of the modified Bessel functions.

Inserting the asymptotical perturbative expansion $S^{(1)}(\tau)$
into equation (\ref{gyeq3}), we calculate functional integral beyond
simple perturbative expansion. We can render our result as a partial
resumation of the simple perturbative series. Moreover, this
procedure allows the parameter $b$ take both positive (anharmonic
oscillator case) and negative (Higgs case). A similar procedure is
familiar in the renormalization group calculations proposed by
Gell-Mann and Low. In Callan -- Symanzik equation for running
coupling constant the $\beta$ function is calculated perturbatively,
however, the equation is solved non-perturbatively.

\vskip 0.3cm {\bf{Acknowledgements}}. This work was supported by
VEGA projects No. 2/3106/2003 (J.B.) and No. 1/025/03 (P.P). The
authors acknowledge discussions with J. Polonyi and M. Znojil and
their interest in this problem.

\end {document}